\documentclass[a4paper]{article}
\pdfoutput=1
 
 \usepackage[margin=1.25in]{geometry}
\usepackage{verbatim}
\usepackage{amssymb,enumerate}
\usepackage{graphicx}
\usepackage{amsbsy}
\usepackage{authblk}
\usepackage{amsmath,amssymb,color}
\usepackage{graphicx}
\usepackage{amsmath,amsfonts}
\usepackage{enumerate}
\usepackage{verbatim,amssymb,amsfonts,pifont,paralist}

\newtheorem{theorem}{Theorem}

\newtheorem{lemma}{Lemma}
\newenvironment{proof}{\paragraph{Proof:}}{\hfill$\square$}

\title{A Note on Plus-Contacts, Rectangular Duals, and Box-Orthogonal Drawings\thanks{Work of the authors is supported in part by 
 NSERC.}}

\author{Therese Biedl$^\S$ and Debajyoti Mondal$^\dagger$} 

\affil{$^\S$Cheriton School of Computer Science, University of Waterloo, Canada\\
	$^\dagger$Department of Computer Science, University of Saskatchewan, Canada\\
  \texttt{biedl@uwaterloo.ca, dmondal@cs.usask.ca}}
  
\usepackage{graphicx,amssymb,amsmath}
\usepackage{tikz}
\usepackage[disable]{todonotes}
\usepackage{paralist}

\title{A Note on Plus-Contacts, Rectangular Duals, and Box-Orthogonal Drawings\thanks{Work of the authors is supported in part by 
 NSERC.}}

\author{
Therese Biedl
        \and
Debajyoti Mondal 
}

\newcommand{\R}{\mathcal{R}}
\newcommand{\plus}{\mathord{\begin{tikzpicture}[baseline=0ex, line width=1, scale=0.13]
\draw (1,0) -- (1,2);
\draw (0,1) -- (2,1);
\end{tikzpicture}}}

\newcommand{\dplus}{\mathord{\begin{tikzpicture}[baseline=0ex, line width=.5, scale=0.13]
\draw (0,0) -- (2,0);
\draw (0,0) -- (1,2);
\draw (2,0) -- (1,2);
\draw (1,0.1) -- (1,1.1);
\draw (0.5,.6) -- (1.5,.6);
\end{tikzpicture}}}

\begin{document}

\maketitle

\begin{abstract}
A plus-contact representation of a planar graph $G$ is called $c$-balanced if for
 every plus shape $\plus_v$, the number of other plus shapes incident to each arm of
 $\plus_v$ is at most $ c \Delta +O(1)$, where $\Delta$ is the maximum 
 degree of $G$. 
 Although small values of
 $c$ have been achieved for a few subclasses of planar graphs (e.g., $2$- and $3$-trees),
 it is unknown whether $c$-balanced representations with
 $c<1$ exist for arbitrary planar graphs.  
 
 In this paper we compute $(1/2)$-balanced plus-contact representations for all planar
 graphs that admit a rectangular dual.
\todo{TB: I shortened the abstract; it restated things that were said already}
  Our result implies that any graph with a rectangular dual has a
 1-bend box-orthogonal drawings such that  for each vertex $v$, 
 the box representing $v$ is a square of side length $\frac{\deg(v)}{2}+ O(1)$.
\end{abstract}

\section{Introduction}
Contact representation of planar graphs have been examined using different geometric objects (e.g., lines, rectangles, triangles, or circles) since the early 1980s~\cite{DuncanGHKK12,Fraysseix91,KobourovMN12,Koebe,Thomassen86}. Besides the intrinsic theoretical interest, such contact layouts find application in applied fields such as cartography, VLSI floor-planning, and data visualization. A rich body of literature examines contact layouts 
 using polygons~\cite{AlamEGKP14,AlamBFGKK13,DuncanGHKK12}, $T$-shapes~\cite{fraysseix_T},  $L$-shapes~\cite{KobourovUV13}, and straight line segments~\cite{Fraysseix91}. 

\begin{figure}[t]
\centering
\includegraphics[width=.8\textwidth]{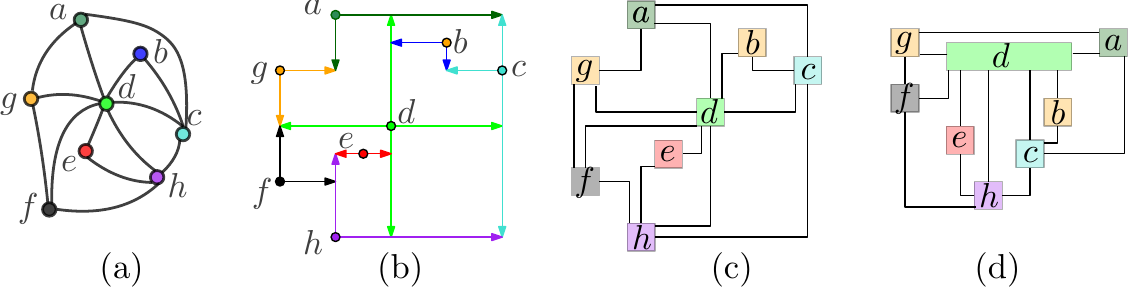} 
\caption{(a) A planar graph $G$. (b)--(c) A $(1/2)$-balanced plus-contact representation of $G$, and a corresponding 1-bend box-orthogonal drawing. 
 (e)  Another box-orthogonal representation of $G$, where vertex $d$ has a large side length. 
} 
\label{fig:pcr}
\end{figure}

 We examine contact representations using \emph{plus shapes}  (i.e., a pair of  intersecting vertical and horizontal line segments). A \emph{plus-contact representation} of an $n$-vertex planar graph $G$ is a non-crossing arrangement $\Gamma_+$ of $n$  plus shapes  such that each vertex $v$ of $G$  is mapped to a distinct plus shape $\plus_v$ in $\Gamma_+$ and two plus shapes in $\Gamma_+$ touch if and only if the corresponding vertices are adjacent in $G$. If no arm of $\plus_v$ is incident to more than $ c\Delta  +O(1)$ other arms, then $\Gamma_+$ is called a \emph{$c$-balanced} representation, e.g., see Fig.~\ref{fig:pcr}(a)--(b). 

Balanced plus-contact representations are motivated by the application of computing 1-bend box-orthogonal drawings with boxes  of small size and constant aspect ratio~\cite{DBLP:journals/algorithmica/PapakostasT00,DBLP:conf/gd/Wood99}, e.g., see Fig.~\ref{fig:pcr}(b)--(d). A \emph{$1$-bend box-orthogonal drawing} (resp., \emph{1-bend Kandinsky drawing (KD)}) is a planar drawing, where each vertex is represented as an axis-aligned box (resp., square) and each edge is represented as an orthogonal polyline (with at most one bend) between the corresponding boxes. Every $c$-balanced plus-contact representation can be transformed into a 1-bend  box-orthogonal drawing with square-size boxes of side length  $c\Delta+O(1)$~\cite{GD13}.  Besides, balanced representations have been useful to construct planar drawings with small number of distinct edge slopes~\cite{GD13,DBLP:conf/gd/GiacomoLM16}. 
 Well balanced representations are known only for $2$-trees  ($1/4\le c\le 1/3$) and planar $3$-trees ($1/3\le c\le 1/2$)~\cite{GD13}. 
 It is not yet known whether there exist $c$-balanced plus-contact representations for arbitrary planar graphs with $c<1$.


We construct $(1/2)$-balanced plus-contact representations of graphs that admit rectangular duals. 
 These graphs are irreducible triangulations (see e.g.~\cite{DBLP:journals/dm/Fusy09}), 
 i.e., graphs where the outer-face has degree at least 4, all inner faces are triangles, and there are no triangles
 that are not face. Our result implies that these graphs have 1-bend box-orthogonal drawings with squares of side 
 length at most $\frac{\deg(v)}{2}+O(1)$ for each vertex $v$.  To our knowledge, this result is new.  The closest
 related results are 2-bend planar drawings where the length of the longer side of the box of $v$ is at most
 $\frac{\deg(v)}{2}+O(1)$~\cite{DBLP:journals/comgeo/BiedlK98}, or 1-bend planar drawing where the length of the
 longer side of the box of $v$ is at most $\deg(v)$ \cite{DBLP:conf/esa/BiedlK97}.   If the planarity requirement
 is dropped, then there are 1-bend orthogonal drawings where the length of the longer side of the box is at most
 $\frac{\deg(v)}{2}+O(1)$ \cite{DBLP:conf/esa/BiedlK97}.

\section{Preliminaries}

\begin{figure}[t]
\centering
\includegraphics[width=.65\textwidth]{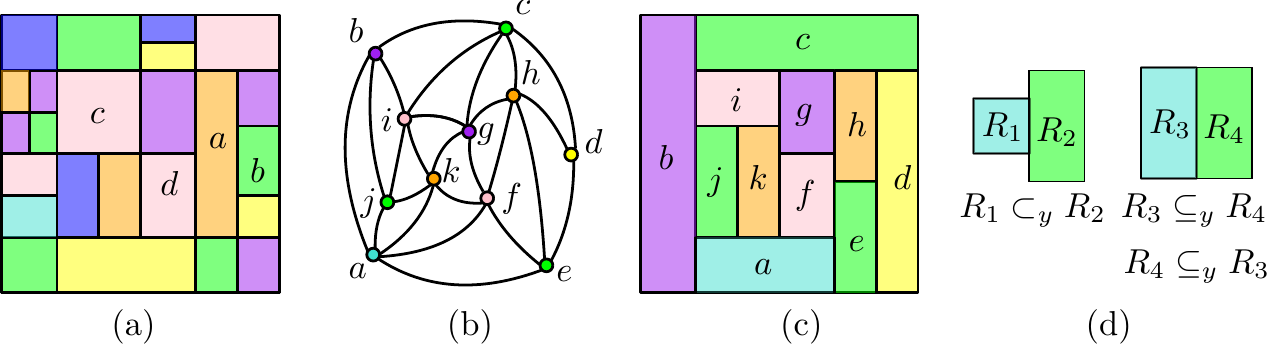}
\caption{(a) A consistent rectangle contact representation. (b) A planar graph $G$.  (c) A rectangular dual of $G$. (d) $R_1\subset_y R_2$, $R_3\subseteq_y R_4$.}
\label{dual}
\end{figure}

Let $R_1$ and $R_2$ be two interior-disjoint rectangle in the plane.
 $R_1$ and $R_2$ are called \emph{adjacent} if they intersect only at their boundaries, i.e., they touch but do not overlap.  If $R_1$ and $R_2$ intersect at a single point 
then we call them \emph{corner adjacent}, e.g., see $c,d$ in Fig.~\ref{dual}(a). On the other hand, if $R_1$ and $R_2$ share a   vertical (horizontal) line segment of non-zero length on their boundaries, then we call them \emph{vertically (horizontally)  adjacent}, e.g., see $a,b$ in Fig.~\ref{dual}(a).

A \emph{rectangular tiling} $\Gamma$ is a partition of a rectangle into non-overlapping rectangles, e.g., see Fig.~\ref{dual}(a) and (c).  This naturally defined a graph $g(\Gamma)$ by assigning one vertex per rectangle and adding edges if and only if the rectangles are adjacent.
We allow four rectangles to meet at a point, which means that $g(\Gamma)$ may be nonplanar, e.g., see Fig.~\ref{basics}(a)--(b) in  Appendix A. 
(Such graphs are also known as \emph{map graphs}.) 
A graph $G$
has a \emph{rectangle contact representation} if there is a rectangular tiling $\Gamma$ with $g(\Gamma)=G$.
\todo{Why did the def'n restate what a tiing is?}
 A \emph{rectangular dual} $R$ of a planar graph $G$ is a rectangle contact representation  with the additional constraint that no four rectangles in $R$ meet at a point. Unlike rectangle contact representations, rectangular duals can exist only for planar graphs.  

 Two adjacent rectangles $R_1$ and $R_2$ in $\Gamma$  are 
 \emph{comparable} if their shared segment coincides with a side of one of these rectangles (see $a,b$ in Fig.~\ref{dual}(a)). Otherwise, we call them \emph{incomparable} (see $h,f$ in Fig.~\ref{dual}(c)). 
We use $R_1\subseteq_y R_2$ (resp., $R_1\subseteq_x R_2$) to denote that $R_1$ and $R_2$ are vertically (resp., horizontally) adjacent, and one side of $R_1$ is a subset of one side of $R_2$, see Fig.~\ref{dual}(d).   
$\Gamma$ is called \emph{consistent}  if   every pair of adjacent rectangles in $\Gamma$ is comparable, see Fig.~\ref{dual}(a).
We create plus-contact representations initially only for consistent rectangle contact representations, and so need
a result whose proof is in Appendix~A.

\begin{lemma}\label{lem:consistent}
For any rectangle contact representation  $\R$
there exists a consistent rectangle contact representation $\R_c$ such that any comparable pair $R_1,R_2$ in $\R$ has the same (vertical or horizontal) adjacency in $\R_c$ along a segment of non-zero length.  $\R_c$ can be found in polynomial time.
\end{lemma}

Let $G$ be a simple and connected planar graph. An \emph{orthogonal drawing} $\Gamma$ of $G$ is a planar drawing of $G$ in $\mathbb{R}^2$, where  each vertex of $G$ is mapped to a point and each edge is mapped to an orthogonal polygonal chain between its corresponding end points. 
We call an orthogonal polygonal chain $P$ a \emph{zigzag path} if it is $x$ or $y$-monotone, and contains at least two bend vertices.  
It is well-known that such zigzags can be eliminated in the following sense.
Two orthogonal drawings $\Gamma,\Gamma'$ of $G$ are \emph{equivalent} if for every vertex $v$ with incident edge $e$, the attachment point of $e$ at $v$ (i.e., east, west, north, south) is the same in $\Gamma$ and $\Gamma'$.  Based on Tamassia's topology-shape-metric approach for
orthogonal drawings, we have:

\begin{lemma} [\cite{TT89a}]
\label{lem:equivalent}
For every  planar orthogonal drawing, there exists an  equivalent planar orthogonal drawing that does not contain any zigzag path. 
\end{lemma}

\section{Drawing Algorithm}

In this section we show that if $G$ admits a rectangular dual $\R$, then it has a $(1/2)$-balanced plus-contact representation $\Gamma_+$. To compute $\Gamma_+$, we first transform $\R$ into a consistent rectangle contact representation $\R_c$ using Lemma~\ref{lem:consistent}, and then transform $\R_c$ into a $(1/2)$-balanced plus-contact representation $\Gamma$ of $g(\R_c)$. Finally, we will modify $\Gamma$ to construct the required 
 representation $\Gamma_+$. 

The representation $\Gamma$ of the supergraph $g(\R_c)$ is already enough
to construct a 1-bend box-orthogonal drawing of $G$ with square-boxes of side length
 $\frac{\deg(v)}{2}+ O(1)$ for every vertex $v$. Therefore, the transformation from
 $\Gamma$ to $\Gamma_+$ (which involves a very large number of cases) is mostly of
theoretical interest, and will be explained in detail only in Appendix B.
 For convenience, we will use the shortcut $\dplus_v:=\frac{\deg(v)}{2}+O(1)$.
 Furthermore, we ignore floors and ceilings as they do not affect the
  asymptotic nature of our results.

\begin{figure}[pb]
\centering
\includegraphics[width=.8\textwidth]{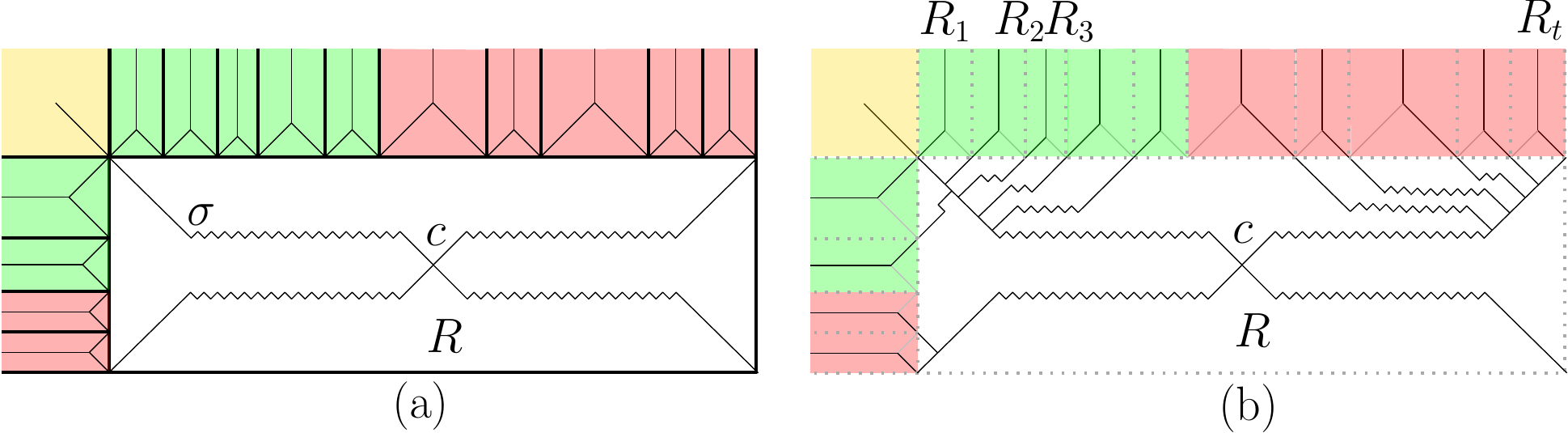}
\caption{(a) Construction of the cords of $R$. (b) Extension of the cords of the rectangles that are adjacent to $R$. Only the rectangles that  lie above or to the left of $R$ are shown.} 
\label{fig:transform2}
\end{figure}

\textbf{From $\R_c$ to $\Gamma$:} Let $v$ be a vertex represented by rectangle $R$ in $\R_c$. 
We first add inside $R$ two polygonal paths $\sigma$ and $\sigma'$ connecting the opposite corners of $R$; see Fig.~\ref{fig:transform2}(a).  These paths are such that after a $45^\circ$-rotation they would be $xy$-monotone orthogonal paths.
At the intersection point $c$ of $\sigma$ and 
$\sigma'$ the path from top-left to bottom-right uses $\diagdown$ while the other path uses $\diagup$.  
\todo{TB: I've removed Appendix B.  Since we're using Tamassia's approach, there is really no need
to give explicit constructions.}
Let the four {\em cords} of $v$ be the four subpaths from $c$ to the corners of $c$.
The crucial insight is that the cords (after a $45^\circ$-rotation) become zig-zag paths, and so all bends can be removed by Lemma~\ref{lem:equivalent}.  Thus this shape is a plus-shape $\plus_v$ with $c$ at the center and the four cords becoming the four arms.


 We now extend the cords of the neighbours of $v$ to realize the required adjacencies. These extensions may add more bends, but we will ensure that the extensions are $xy$-monotone paths (after a rotation) that begin and end with the same type of diagonal. Hence these are again  zig-zag paths and all bends can be removed to obtain a plus-contact representation.
 We must ensure that at most $\dplus_v$ contacts are on each cord of $v$.
%
 Let $R_1,\ldots,R_t$ be the rectangles from left to right that are incident to the top boundary of $R$.
We know that $R\subseteq_x R_1$ or $R_1\subseteq_x R$ since we have a consistent rectangle contact 
representation.  If $R\subset_x R_1$ then a contact representing this edge will be created inside $R_1$,
not inside $R$.  So assume that $R_i\subseteq_x R$ for all $1\leq i\leq t$.
   We choose the bottom-left cords of the first $t/2$ rectangles
  to touch the top-left cord of $R$, and the bottom-right cords of the remaining rectangles
  to touch the top-right cord of $R$ 
 using zigzag paths, as illustrated in  Fig.~\ref{fig:transform2}(b).
 The treatment for the other sides is symmetric. 
 The top-left cord of $R$ now has 
 $\frac{\delta_t}{2}+\frac{\delta_l}{2}+O(1) \le \dplus_v$ contacts, where $\delta_t$ and 
 $\delta_l$ are the number of rectangles incident to the top and left sides of $R$, respectively.
Similarly all other cords have at most $\dplus_v$ contacts as desired.


 Let the drawing determined by the cords of $R$ be $H$, which 
 we refer to as a \emph{pseudo-plus representation}. 
Now rotate $H$ by  $45^\circ$ to turn all cords into orthogonal
$xy$-monotone paths.
By Lemma~\ref{lem:equivalent}, there exists an 
 equivalent orthogonal drawing $H'$ that contains no zigzag paths,
which means that cords become straight-line segments, hence arms, and
 $H'$ is the required plus-contact representation $\Gamma$. 

\textbf{Time Complexity:} 
 A rectangular dual $\R$ of $G$ can be computed in polynomial time (if it exists)~\cite{BhaskerS88,KozminskiK85}.
 By Lemma~\ref{lem:consistent}, $\R$ can be transformed  into $\R_c$ in polynomial time. 
 Consider now the construction of 
the pseudo-plus representation $H$. 
 The time complexity for this may initially appear high since cords may have many bends.
 However, instead of computing the
	pseudo-plus representation $H$ explicitly, we only describe it
 implicitly via the topology-shape metric approach introduced by Tamassia~\cite{Tamassia87}.
This will lead to overall polynomial time.
 
\begin{figure}[pt]
\centering
\includegraphics[width=\textwidth]{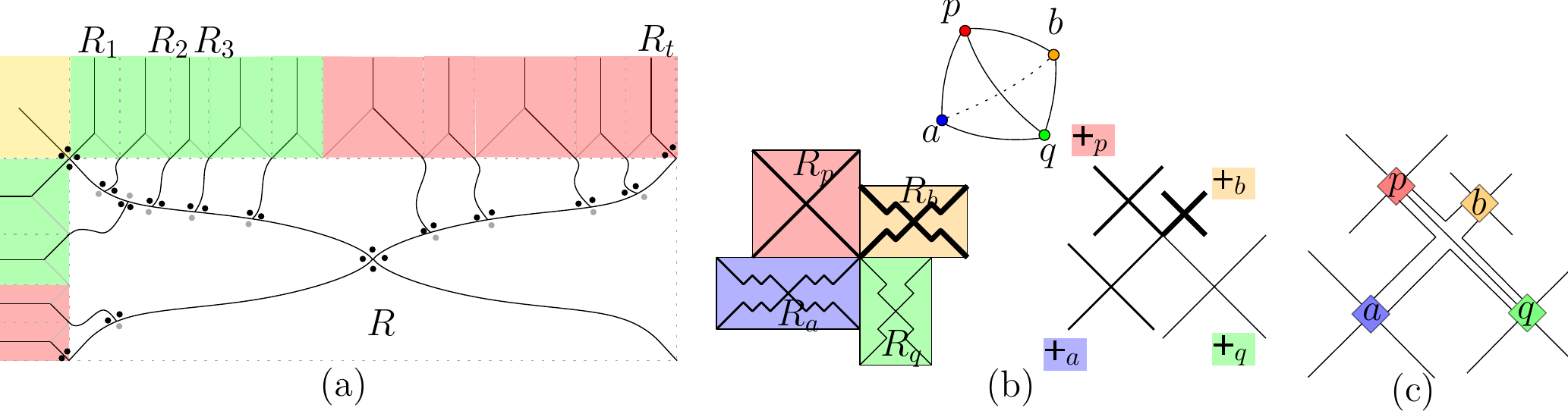}
\caption{(a) Construction of the cords of $R$. The $90^\circ$ and 
 $180^\circ$ angles are marked in black and gray dots, respectively. (b) Transformation into 1-bend Box orthogonal drawings.}
\label{fig:x}
\end{figure}

Specifically, $H$ can be described by defining a graph whose vertices are the
\todo{This part got rewritten a lot.}
ends of the cords, and whose edges are the parts of the cords between ends
or contact points.  At every
center, the four incident face-angles are $90^\circ$.  At every touching point
we have one cord touching the interior of another, which gives two incident
face-angles of $90^\circ$ and one of $180^\circ$.  At every end of a cord,
we have some number of cords ending at the same point, but again, the
incident face-angles are prescribed by our construction.
Hence we know all face-angles at vertices.  We also know that for any edge there exists
a drawing such that the {\em bend-number} (defined to be the difference
between left turns and right turns when walking from one end to the other)
is 0.  Since there exists an orthogonal drawing that respects these
face-angles and bend-numbers, one can use the approach of 
Tamassia~\cite{Tamassia87} to find an orthogonal drawing $H'$ that realizes
the face-angles and bend-numbers and has no zig-zags in polynomial time.
This is the desired plus-contact representation.
 

\textbf{From $\Gamma$ to a $1$-Bend Box-orthogonal drawing:}
Now convert $\Gamma$ into a $1$-bend box-orthogonal drawing
as explained in \cite{GD13}.  Briefly, this places a box for $v$
at the center of $\plus_v$ and routes the edges along the arms
of $\plus_v$; with some offset to avoid overlap.
 Observe that $G$ is a subgraph of $g(\R_c)$.  
  Every four mutually adjacent rectangles $\{R_a,R_p,R_b,R_q\}\in \R_c$, 
  give rise to exactly one adjacency in $g(\R_c)$ that is 
  not in $G$, e.g., see Fig.~\ref{fig:x}(b).
 This undesired adjacency also appears in the 
  plus-contact representation $\Gamma$ and consequently in the
  $1$-bend box-orthogonal drawing. However, we can simply remove
 this edge from the drawing,
 as illustrated in Fig.~\ref{fig:x}(c), and obtain:

\begin{theorem}
Let $G$ be a planar graph that admits a rectangular dual. Then 
$G$ has  a 1-bend box-orthogonal drawing, where each
 vertex $v$ is a square of side length at most $\frac{\deg(v)}{2}+ O(1)$.
\end{theorem}

{\textbf{From $\Gamma$ to $\Gamma_+$:}} 
We would like to transform $\Gamma$ to remove unnecessary adjacencies.
Actually, we will
modify the pseudo-plus representation $H$ instead, since the changes require extending
cords in different ways.  The resulting pseudo-plus representation $H^+$ of $G$
can be transformed into $\Gamma_+$ as before.

\begin{figure}[t]
\centering
\includegraphics[width=\textwidth]{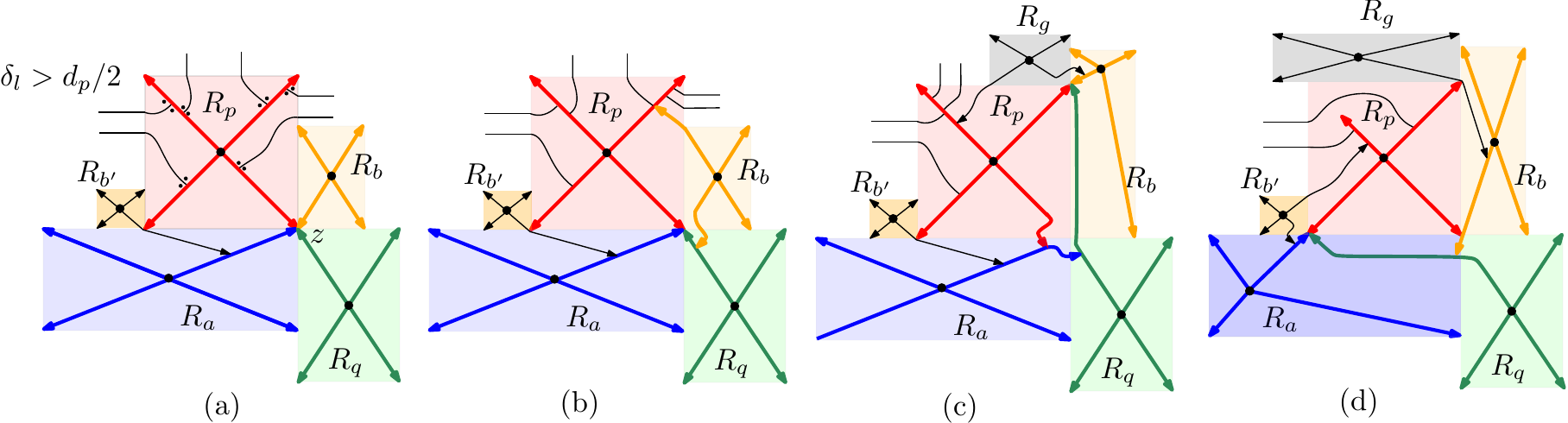}
\caption{Illustration for Case 1a, where $\delta_l > d_p /2$. (a) Schematic representation of the initial configuration. The plus shapes are drawn with bidirected lines. The thin lines represent the distribution of $\delta_l,\delta_r,\delta_t,\delta_b$ to the four cords of $R$. The black dots represent  $90^\circ$ angles. 
(b-d) Resolution for various sub-cases.}
\label{case1}
\end{figure}

Consider Fig.~\ref{case1}(a).
 Let  $R_a,R_p,R_b,R_q$ be four  mutually adjacent rectangles in $\R_c$,
 in this clockwise order around  their common corner $z$ and starting
 with the bottom-left rectangle.
 One of the edges $(a,b)$ or $(p,q)$  did not exist in $G$, say
 $(a,b)$ was unnecessary.  We refer to $R_a,R_b$ as an \emph{excess pair},
 and $R_p$ as the \emph{consumer} of this excess pair.
Put differently, the consumer of an unnecessary edge $(a,b)$
is the upper one of the two rectangles that share a corner with $R_a$ and $R_b$.
We re-route locally near the
consumer $R_p$ 
%
such that
 (A) all unnecessary  adjacencies for which $R_p$ is the consumer have been removed,
 (B) all other adjacencies within the neighbours of $R_p$ have been retained,
 (C) no new unnecessary adjacency is introduced, 
 (D) all cords remain $xy$-monotone paths (after a 45$^\circ$ rotation),
 and
 (E) the cords of $R_p$ remain $(1/2)$-balanced. 

The details of processing a consumer $R_p$ are unfortunately quite tedious;
Fig.~\ref{case1} shows three of the (many) cases and Appendix B gives full details. 
Applying this to all consumers  gives a pseudo-plus representation of $G$, which
 can be transformed to $\Gamma_+$, and we obtain:

\begin{theorem}
Let $G$ be a planar graph that admits a rectangular dual. Then  
 $G$ has a plus-contact representation where for each vertex $v$
 each arm of $\plus_v$ has at most $\frac{\deg(v)}{2}+O(1)$ contacts with other plus-shapes.  
\end{theorem}


\section{Conclusion}
We have shown that every planar graph with a rectangular dual has a $(1/2)$-balanced plus-contact representation and a 1-bend box-orthogonal drawing with square-size boxes of side length  $\frac{\deg(v)}{2}+O(1)$ (for each vertex $v$).    Both representations can be computed in polynomial time.
While our results hold for all 4-connected planar graphs with four outer vertices,  it remains open whether every planar graph admits a $c$-balanced representation for some $c<1$. 


 
\bibliography{ref,bibs}

\begin{thebibliography}{10}

\bibitem{AlamBFGKK13}
M.~J. Alam, T.~Biedl, S.~Felsner, A.~Gerasch, M.~Kaufmann, and S.~G. Kobourov.
\newblock Linear-time algorithms for hole-free rectilinear proportional contact
  graph representations.
\newblock {\em Algorithmica}, 67(1):3--22, 2013.

\bibitem{AlamEGKP14}
M.~J. Alam, D.~Eppstein, M.~T. Goodrich, S.~G. Kobourov, and S.~Pupyrev.
\newblock Balanced circle packings for planar graphs.
\newblock In {\em Proceedings of the 22nd International Symposium on Graph
  Drawing (GD)}, volume 8871 of {\em LNCS}, pages 125--136. Springer, 2014.

\bibitem{BhaskerS88}
J.~Bhasker and S.~Sahni.
\newblock A linear algorithm to find a rectangular dual of a planar
  triangulated graph.
\newblock {\em Algorithmica}, 3:247--278, 1988.

\bibitem{DBLP:journals/comgeo/BiedlK98}
T.~Biedl and G.~Kant.
\newblock A better heuristic for orthogonal graph drawings.
\newblock {\em Computational Geometry}, 9(3):159--180, 1998.

\bibitem{DBLP:conf/esa/BiedlK97}
T.~Biedl and M.~Kaufmann.
\newblock Area-efficient static and incremental graph drawings.
\newblock In {\em Proceedings of the 5th Annual European Symposium on
  Algorithms (ESA)}, volume 1284 of {\em LNCS}, pages 37--52. Springer, 1997.

\bibitem{BiedlLPS13}
T.~Biedl, A.~Lubiw, M.~Petrick, and M.~J. Spriggs.
\newblock Morphing orthogonal planar graph drawings.
\newblock {\em {ACM} Transactions on Algorithms}, 9(4):29, 2013.

\bibitem{Fraysseix91}
H.~de~Fraysseix, P.~O. de~Mendez, and J.~Pach.
\newblock Representation of planar graphs by segments.
\newblock {\em Intuitive Geometry}, 63:109--117, 1991.

\bibitem{fraysseix_T}
H.~de~Fraysseix, P.~O. de~Mendez, and P.~Rosenstiehl.
\newblock On triangle contact graphs.
\newblock {\em Combinatorics, Probability and Computing}, 3(2):233--246, 1994.

\bibitem{DuncanGHKK12}
C.~A. Duncan, E.~R. Gansner, Y.~F. Hu, M.~Kaufmann, and S.~G. Kobourov.
\newblock Optimal polygonal representation of planar graphs.
\newblock {\em Algorithmica}, 63(3):672--691, 2012.

\bibitem{GD13}
S.~Durocher and D.~Mondal.
\newblock On balanced $\plus$-contact representations.
\newblock In {\em Proceedings of the 21st International Symposium on Graph
  Drawing (GD)}, volume 8242, pages 143--154. Springer, 2013.

\bibitem{DBLP:journals/dm/Fusy09}
{\'{E}}.~Fusy.
\newblock Transversal structures on triangulations: {A} combinatorial study and
  straight-line drawings.
\newblock {\em Discrete Mathematics}, 309(7):1870--1894, 2009.

\bibitem{DBLP:conf/gd/GiacomoLM16}
E.~D. Giacomo, G.~Liotta, and F.~Montecchiani.
\newblock 1-bend upward planar drawings of {SP}-digraphs.
\newblock In Y.~Hu and M.~N{\"{o}}llenburg, editors, {\em Proceedings of the
  24th International Symposium on Graph Drawing and Network Visualization
  (GD)}, volume 9801 of {\em LNCS}, pages 123--130. Springer, 2016.

\bibitem{KobourovMN12}
S.~G. Kobourov, D.~Mondal, and R.~I. Nishat.
\newblock Touching triangle representations for 3-connected planar graphs.
\newblock In {\em Proceedings of the 20th International Symposium on Graph
  Drawing (GD)}, volume 7704, pages 199--210. Springer, 2012.

\bibitem{KobourovUV13}
S.~G. Kobourov, T.~Ueckerdt, and K.~Verbeek.
\newblock Combinatorial and geometric properties of planar {L}aman graphs.
\newblock In {\em Proceedings of the 24th Annual ACM-SIAM Symposium on Discrete
  Algorithms (SODA)}, pages 1668--1678. SIAM, 2013.

\bibitem{Koebe}
P.~Koebe.
\newblock Kontaktprobleme der konformen {A}bbildung.
\newblock {\em Ber. S\"{a}chs. Akad. Wiss. Leipzig, Math.-Phys. Kl.},
  88:141--164, 1936.

\bibitem{KozminskiK85}
K.~Kozminski and E.~Kinnen.
\newblock Rectangular duals of planar graphs.
\newblock {\em Networks}, 15(2):145--157, 1985.

\bibitem{DBLP:journals/algorithmica/PapakostasT00}
A.~Papakostas and I.~G. Tollis.
\newblock Efficient orthogonal drawings of high degree graphs.
\newblock {\em Algorithmica}, 26(1):100--125, 2000.

\bibitem{Tamassia87}
R.~Tamassia.
\newblock On embedding a graph in the grid with the minimum number of bends.
\newblock {\em {SIAM} Journal on Computing}, 16(3):421--444, 1987.

\bibitem{TT89a}
R.~Tamassia and I.~Tollis.
\newblock Planar grid embedding in linear time.
\newblock {\em {IEEE} Transactions on Circuits and Systems}, 36(9):1230--1234,
  1989.

\bibitem{Thomassen86}
C.~Thomassen.
\newblock Interval representations of planar graphs.
\newblock {\em J. Comb. Theory, Ser. B}, 40(1):9--20, 1986.

\bibitem{DBLP:conf/gd/Wood99}
D.~R. Wood.
\newblock Multi-dimensional orthogonal graph drawing with small boxes.
\newblock In {\em Proceedings of the 7th International Symposium on Graph
  Drawing (GD)}, volume 1731 of {\em LNCS}, pages 311--322. Springer, 1999.

\end{thebibliography}


\begin{appendix}
\newpage
\section*{Appendix A}

\begin{figure}[b]
\centering
\includegraphics[width=.4\textwidth]{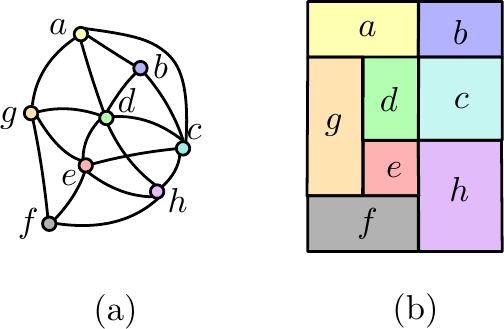}
\caption{
 (a) A graph $G$. (b) A rectangle contact representation of $G$.}
\label{basics}
\end{figure}

\noindent\textbf{Lemma~\ref{lem:consistent}}\emph{
For any rectangle contact representation  $\R$
there exists a consistent rectangle contact representation $\R_c$ such that any comparable pair $R_1,R_2$ in $\R$ has the same (vertical or horizontal) adjacency in $\R_c$ along a segment of non-zero length.  $\R_c$ can be found in polynomial time.
}

\begin{proof}
The idea is to process the incomparable pairs of $\R$ one after another, and at each step ensuring that no adjacency in $\R$ is destroyed.  Every rectangle contact representation can be transformed into an equivalent grid representation, i.e., when the endpoints of all the line segments have integral coordinates.  Therefore, we may assume that $\R$ is a grid representation.  

Here we describe how to remove an incomparable pair that is vertically adjacent. The treatment for the horizontally adjacent incomparable pairs is symmetric. Let $R_p$ and $R_q$ be a pair of vertically adjacent rectangles, which are incomparable. Let $ab$ be the common vertical segment on the boundary of $R_p$ and $R_q$. Without loss of generality assume that $R_p$ lies to the left of $R_q$, and $a$ and $b$ are the top-left and bottom-right corners of $R_q$ and $R_p$, respectively. Fig.~\ref{consistent} 
illustrates such a scenario.

\begin{figure}[pt]
\centering
\includegraphics[width=\textwidth]{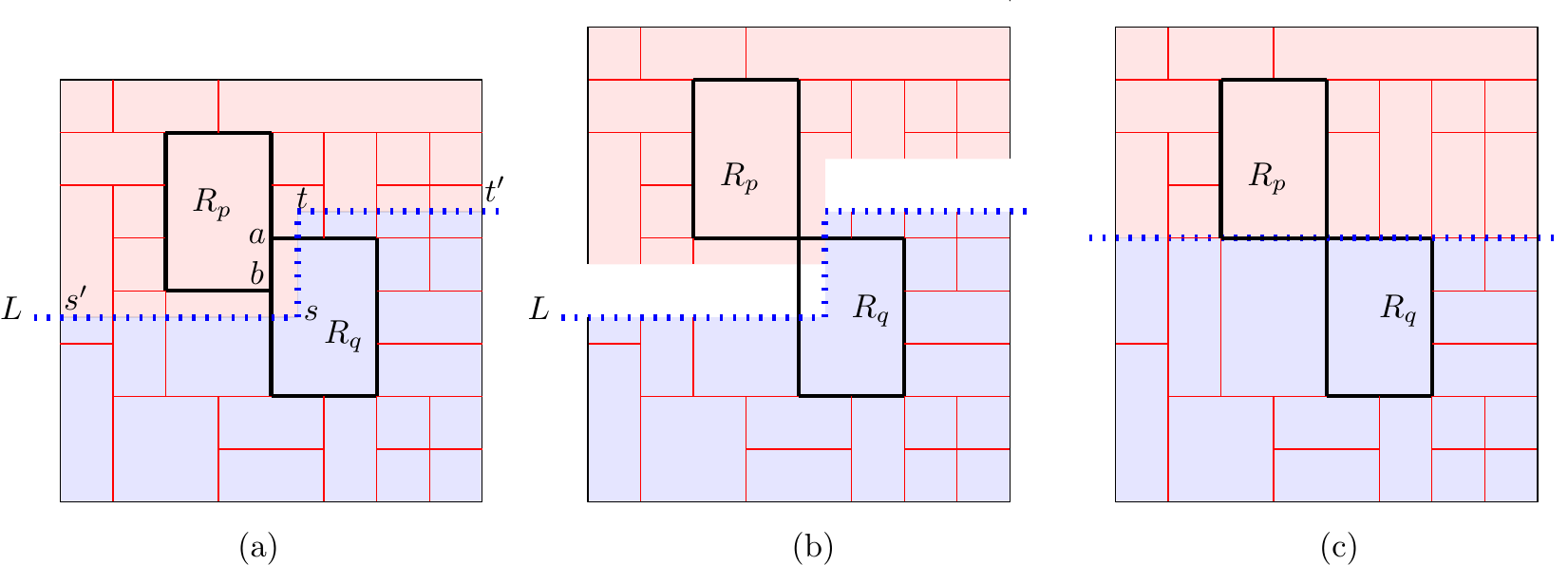}
\caption{(a) A rectangle contact representation, which is not consistent, and a cut corresponding to $ab$.  (b)--(c) Removal of segment $ab$. }
\label{consistent}
\end{figure}

We now modify $\R$ such that the segment $ab$ becomes degenerate, and $R_p$ and $R_q$ become corner adjacent. We first define a cut  that partitions $\R$ into two smaller drawings, as follows. 
 Let $s$ and $t$ be two points  with coordinates $(x(b) + \alpha,  y(b)-\alpha)$ and $(x(a) + \alpha,  y(a)+\alpha)$, respectively, for some constant $\alpha$, where $0<\alpha<1$. Then the cut is an    orthogonal polygonal chain $L=(s',s,t,t')$, where $s'$ and $t'$ lie on the left and right boundary of $\R$, respectively. See Fig.~\ref{consistent}(a).  The notion of cut has previously been   used in the literature in more generalized settings, e.g., in the context of morphing orthogonal drawings~\cite{BiedlLPS13}.
 
Let $\Gamma_t$ be the drawing that consists of all the points of $\Gamma$ lying above $L$. We move all the points of $\Gamma_t$ upward by $|ab|$ units, except the points that lie on the boundary of $R_q$. Consequently, the segment $ab$ becomes degenerate. Let $R_t$ be the rectangle that contains the point $t$.  Observe that  if $L$ intersects the left boundary of a rectangle in $R\not \in \{R_t,R_q\}$, then it also intersects its right boundary at the same height. Hence each of these rectangles can be recovered by extending the left and right boundaries vertically until they reach $L$. Since we do not split the bottom side of $R_t$, we can recover $R_t$ by extending only its right boundary.  Since the length of segment $|ab|$ is integral, all the line segments of the resulting drawing have integral coordinates.   


It is straightforward to remove an incomparable pair in $O(n)$ time, where $n$ is the number of rectangles in $\R$.
 Since $\R$ may contain at most $O(n)$ incomparable pairs, one can construct $\R_c$ in $O(n^2)$ time.
\end{proof}

\todo{TB: I've removed the paragraph of the details of how to extend cords; this
really wasn't saying much that wasn't in the main text.}

\section*{Appendix B}

Here we describe the details of processing a consumer $R_p$.  
 Let $p$ be the vertex that corresponds to $R_p$. Let $d_p$ denote
 the degree of $p$, and let $\delta_t,\delta_b,\delta_l,\delta_r$ 
 be the number of rectangles that are incident to the top, bottom, left and right
 sides of $R_p$, respectively. Let $\Psi_{tl},\Psi_{tr},\Psi_{bl},\Psi_{br}$
 be the number of contact points on the top-left, top-right, bottom-left, and bottom-right
 cords of $R_p$. 

We have numerous cases, depending on whether $R_p$ is the consumer of one excess pair
or of two.  We must further distinguish by whether certain neighbours of $R_p$ contain
the $x$-range resp.~$y$-range of $R_P$.  (We will not always explicitly say in the text
which neighbour is meant when speaking, e.g., of $R_g$; this should be clear from the
picture.)  Finally we distinguish by the size and
relationships between $\delta_t,\delta_b,\delta_l,\delta_r$.  Unfortunately there
appears to be no way to unify these cases into fewer.  However, the following 
observation will often be used to argue correctness.  Assume that we are in a setup where
we can {\em saturate} one cord, i.e., add contacts such that (e.g.) $\psi_{tl}=\delta_P/2$.
Then, as long as we assign all other contacts at $P$ to other cords, all cords have 
at most $\delta_P/2$ contacts as required.

\noindent\textbf{Case 1 ($R_p$ is a consumer of exactly one excess pair).} Without loss of generality 
 assume that the excess pair $\{R_a, R_b\}$ is at the bottom-right corner of $R_p$,
 and $R_b$ lies above $R_a$, e.g., see Fig.~\ref{case1}(a). 
 
\noindent\textbf{Case 1a ($R_p\subset_x R_a$).} This case is illustrated in Fig.~\ref{case1}(a).
  Here we distinguish two scenarios depending on whether $\delta_l > d_p /2$ or $\delta_l \le d_p /2$. 


 \begin{itemize}
\item Case ($\delta_l > d_p /2$):  This implies that $\delta_r\le d_p/2$.  
 \begin{itemize}
\item 
 If $R_b\subseteq_y R_p$, then we re-route the cords of $R_p$ following Fig.~\ref{case1}(b)
and saturate the bottom-left cord.
%
We note that in Fig.~\ref{case1}, we have $R_{b'}\subset_x R_a$, which is merely an illustration. The  modification works fine even when $R_{a}\subseteq_x R_{b'}$. The same applies to all the adjacencies that do  not involve $R_p$. 

 \item
 If $R_p\subset_y R_b$ and $R_g\subseteq_x R_p$,
 then we follow Fig.~\ref{case1}(c) and saturate the bottom-left cord.

 Observe that if $R_g$ is a consumer, then the bottom-left corner of $R_g$ will 
 coincide with the top-left corner of $R_p$. This modification removes the excess pair from 
 the bottom corners of $R_p$, but $R_g$ still remains a consumer. The excess pair at the bottom corners
 of $R_g$ will be removed when we process $R_g$. 
 \item 
 If $R_p\subset_y R_b$ and $R_p\subset_x R_g$,
 then we follow Fig.~\ref{case1}(d) and saturate the top-left cord.
 Here the top-left cord of $R_p$ cannot reach the top-left corner  of $R_p$, which is fine since the adjacency between  $R_g$ and $R_p$ is realized at the top-right corner of $R_p$, and since for each rectangle adjacent to the left of $R_p$, one of its two cords is extended to touch the cords of $R_p$. 
 We will never need to choose between the top-left and bottom-left arms 
 of $R_p$ to process the remaining consumer rectangles.
 \end{itemize}
 \end{itemize}

 \begin{itemize}
 \item Case ($\delta_l \le d_p /2$ and  $R_{b'}\subseteq_y R_p$):
 \begin{itemize} 
 \item If $\delta_r > d_p /2$, then we follow Fig.~\ref{case1b}(b)
and saturate the bottom-right cord.
 
 \item If $\delta_r \le d_p /2$ and $R_b\subseteq_y R_p$,
 then we follow Fig.~\ref{case1b}(c)--(d) depending on whether 
 $\delta_r \le \delta_l$ or not. 

If $\delta_r \le \delta_l$ (Fig.~\ref{case1b}(c)),
 then  $\Psi_{tl} \le \frac{\delta_t}{2} +O(1)\le 	\dplus_p$, and 
  $\Psi_{tr} \le \frac{\delta_t}{2} +\delta_r +O(1)\le \frac{\delta_t+ 2\delta_r}{2} +O(1)
 \le  \frac{\delta_t+ \delta_l+\delta_r}{2} +O(1)\le 	\dplus_p$.
  If $\delta_r > \delta_l$ (Fig.~\ref{case1b}(d)), then 
  $\Psi_{tr} \le \frac{\delta_t}{2} +\frac{\delta_r}{2} +O(1)\le 	\dplus_p$,
  and 
  $\Psi_{tl} \le \frac{\delta_t}{2} +\delta_l +O(1)\le \frac{\delta_t+ 2\delta_l}{2} +O(1)
 \le  \frac{\delta_t+ \delta_l+\delta_r}{2} +O(1)\le 	\dplus_p$.

  
\item If $\delta_r \le d_p /2$ and $R_p\subseteq_y R_b$, then  the modification is
 shown in Fig.~\ref{case1b}(e). Note that $\delta_l \le d_p /2$. Therefore, 
 we can saturate $\Psi_{tr}$ with $\min\{\delta_t,d_p /2\}$ contacts to have 
 $\Psi_{tl} \le \delta_l+\delta_t- \Psi_{tr} +O(1) \le \dplus_p$. 
 \end{itemize}
 
\begin{figure}[ht]
\centering
\includegraphics[width=\textwidth]{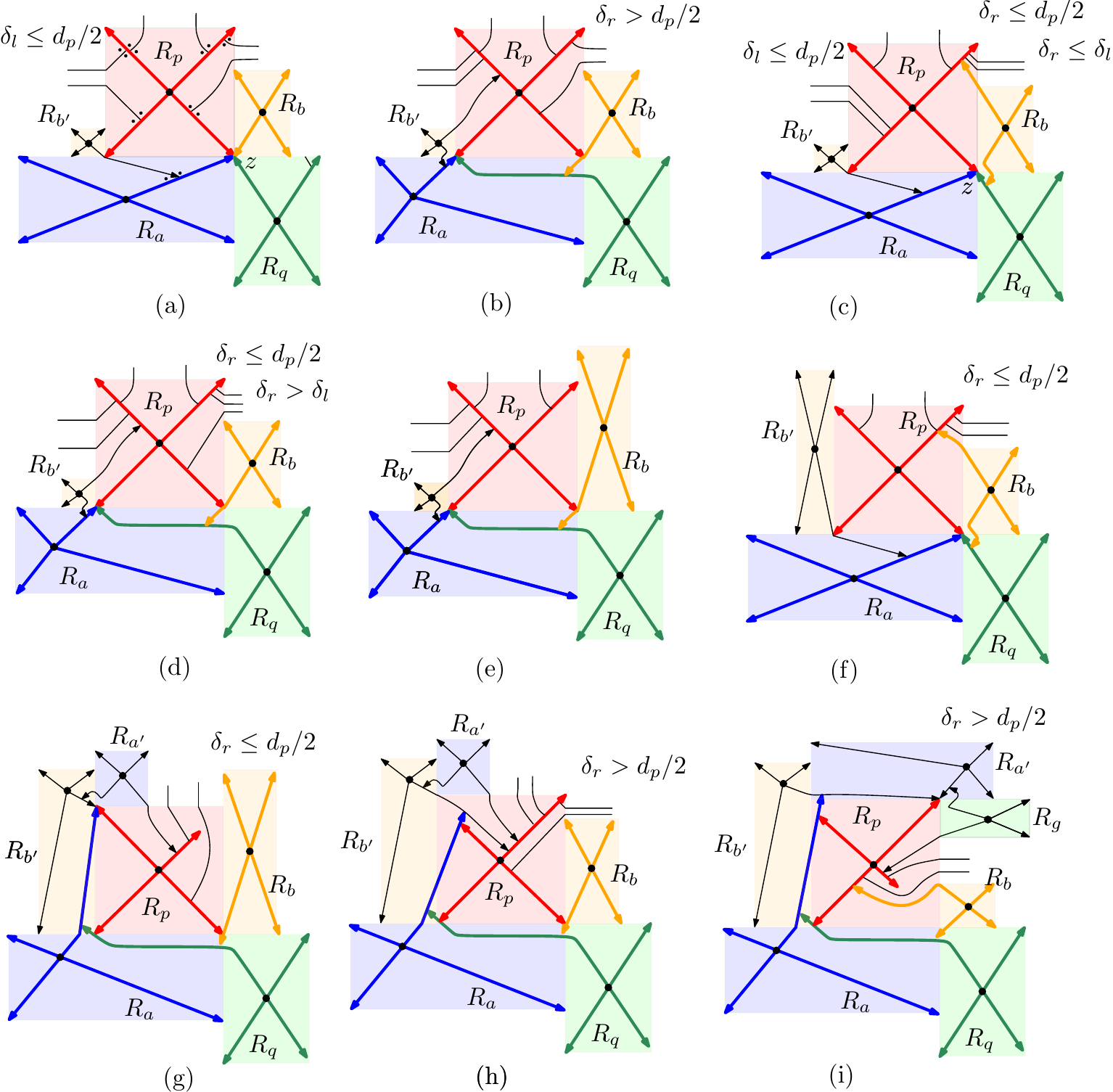}
\caption{Illustration for Case 1a, where $\delta_l \le d_p /2$. (a) Initial configuration. 
(b-i) Resolution.}
\label{case1b}
\end{figure}
 
 \item Case ($\delta_l \le d_p /2$ and  $R_p\subset_y R_{b'}$):
 \begin{itemize} 
 \item If $\delta_r \le d_p /2$ and  $R_b\subseteq_y R_p$,
 then we follow  Fig.~\ref{case1b}(f).
 We can saturate the top-right cord with 
 $d_p/2$ incidences using edges from the top and the right.

\item If $\delta_r \le d_p /2$ and  $R_p\subset_y R_b$,
 then we  follow  Fig.~\ref{case1b}(g).
 Here $\Psi_{tr}, \Psi_{br}\le \delta_t/2 \le \dplus_p$.
Note that  the top-right cord of $R_p$ cannot reach the top-right corner  of $R_p$, which is fine since the adjacency between  $R_b$ and $R_p$ is realized at the bottom-right corner of $R_p$, and since for each rectangle adjacent to the top of $R_p$, one of its two cords is extended to touch the cords of $R_p$. 
 We will never need to choose between the top arms (similarly, bottom arms)  
 of $R_p$ to process  the remaining consumer  rectangles.
 
\item  If $\delta_r > d_p /2$ then we will always saturate the bottom-right cord.
If $R_{a'} \subseteq_x R_p$, then we follow  Fig.~\ref{case1b}(h);  
and if  $R_p \subset_x R_{a'}$, then we follow Fig.~\ref{case1b}(i).  
 We can distribute $\delta_r$ contacts such that $\Psi_{tr},\Psi_{br}, \Psi_{tl}\le \dplus_p$.    
In the latter case the bottom-right cord of $R_p$ cannot reach the bottom-right corner  of $R_p$, which is fine since the  adjacency between  $R_b$ and $R_p$ is realized at the bottom-left cord of $R_p$, and since for each rectangle adjacent to the right of $R_p$, one of its two cords is extended to touch the cords of $R_p$. 
  We will never need to choose between the right arms 
 of $R_p$ to process the remaining consumer rectangles.

   \end{itemize}
 \end{itemize}



\begin{figure}[ht]
\centering
\includegraphics[width=.8\textwidth]{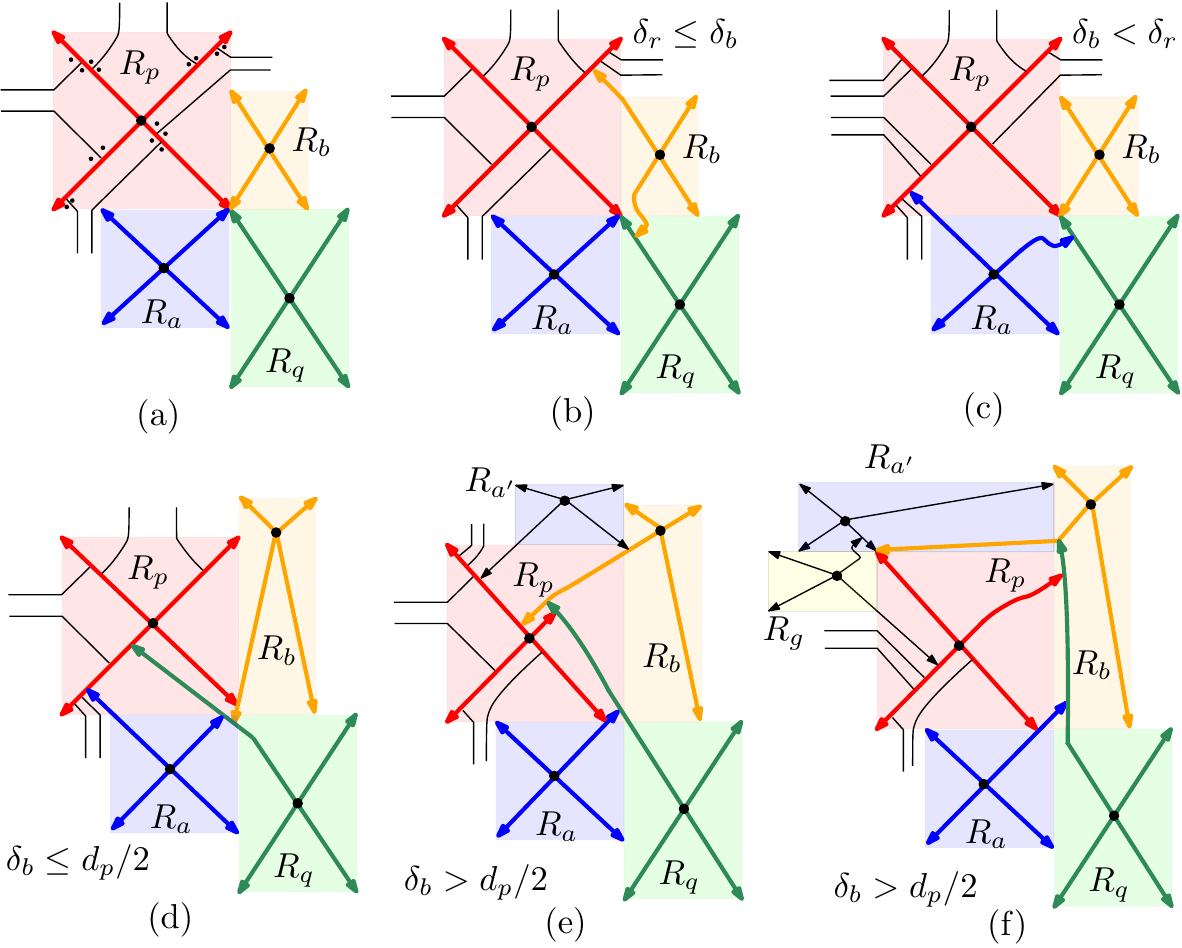}
\caption{Illustration for Case 1b. (a) Initial configuration. 
(b-f) Resolution.}
\label{case1c}
\end{figure}

\noindent\textbf{Case 1b ($R_a\subseteq_x R_p$).} This case is illustrated in Fig.~\ref{case1c}(a).
 Here we distinguish the scenarios whether $R_b\subseteq_y R_p$ or $R_p\subset_y R_b$.

 \begin{itemize}
 \item Case ($R_b\subseteq_y R_p$):

If $\delta_r\le \delta_b$, then
 we follow Fig.~\ref{case1c}(b). It is straightforward
 to see that $\Psi_{bl}, \Psi_{tl} \le \dplus_p$ and $\Psi_{br} \in O(1)$.
 Since $\delta_r\le \delta_b$,      
 $\Psi_{tr}   \le \frac{\delta_t}{2} + \delta_r \le \frac{\delta_t+2\delta_r}{2} \le 
  \frac{\delta_t+\delta_l+ \delta_r}{2} \le d_p /2$. 
 If $\delta_b<\delta_r$, then we follow Fig.~\ref{case1c}(c).
 The analysis for the contact points is symmetric. 
 
 \item Case ($R_p\subset_y R_b$):

If $\delta_b\le \delta_p/2$, then
 we  follow Fig.~\ref{case1c}(d) and can saturate the top-left
cord since $\delta_\ell +\delta_t\geq \delta_p/2 - O(1)$.  On the other hand, if $\delta_b> \delta_p/2$, then we 
 distinguish between whether $R_{a'} \subseteq_x R_p$ or not. 
If $R_{a'} \subseteq_x R_p$, then we follow Fig.~\ref{case1c}(e) and saturate the bottom-right cord.
If $R_p \subseteq_x R_{a'}$, then we follow Fig.~\ref{case1c}(d) and again saturate the bottom-right cord.
\end{itemize}

\begin{figure}[ht]
\centering
\includegraphics[width=\textwidth]{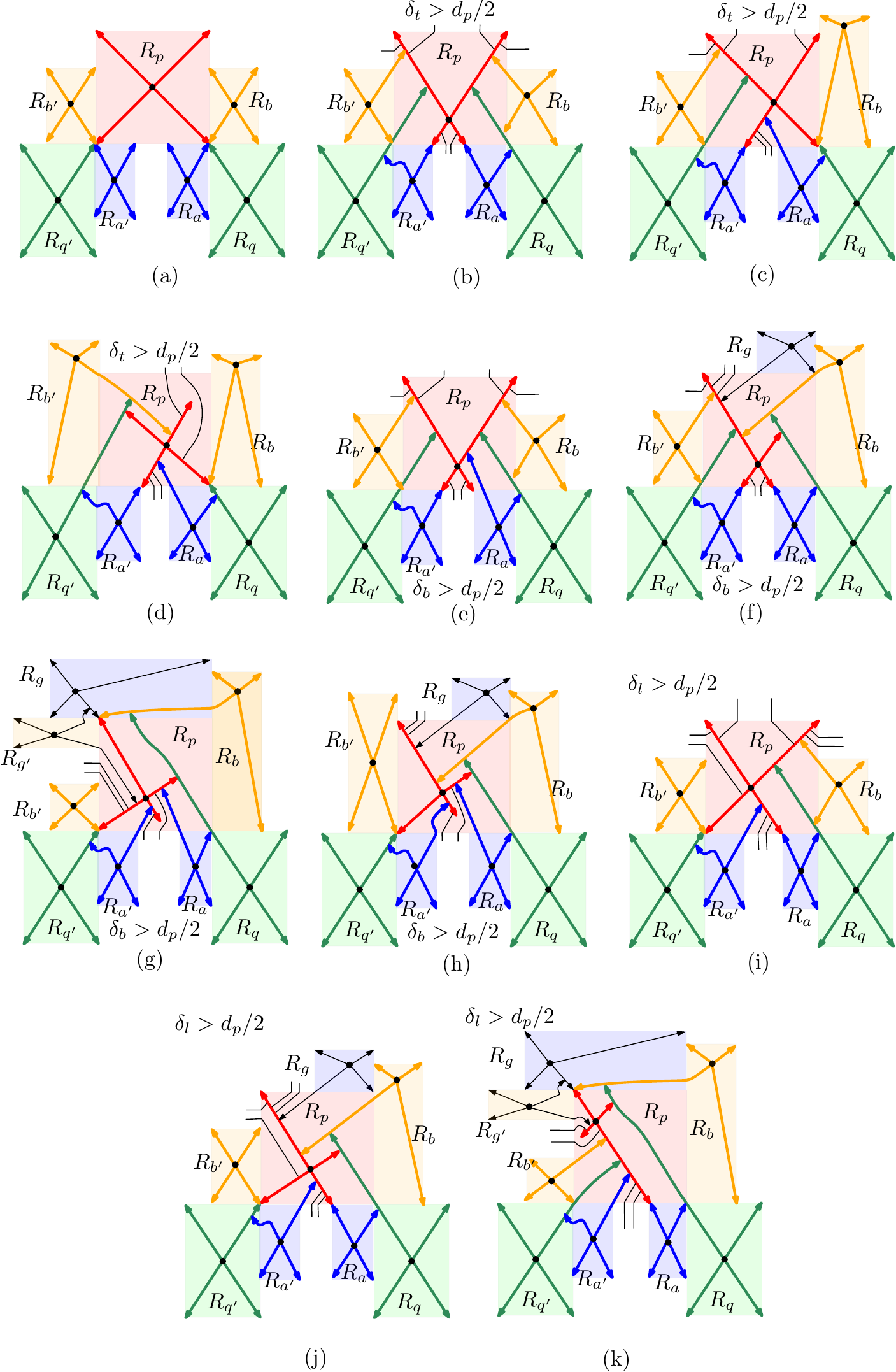}
\caption{Illustration for Case 2. (a) Initial configuration. 
(b-k) Resolution.}
\label{case2}
\end{figure}
 
\noindent\paragraph{Case 2 ($R_p$ is a consumer of two excess pairs).} 
 See Fig.~\ref{case2}(a).
 We consider three cases depending on whether any of  $\delta_l, \delta_r, \delta_t$ or $\delta_b$  is larger than $d_p /2$ or not.

  \begin{itemize} 
 \item Case ($\delta_t > d_p /2$):

In this case, we will always saturate the top-right cord.
If $R_{b'},R_b\subseteq_y R_p$, then we follow Fig.~\ref{case2}(b);
if $R_p\subset_y R_b$ and $R_{b'} \subseteq_y R_p$, then we follow  Fig.~\ref{case2}(c),
and the case when $R_p\subset_y R_{b'}$ and $R_b\subseteq_y R_p$ is symmetric.  
Finally, if   $R_p\subset_y R_b$ and $R_p\subset_y R_b$, then we follow   Fig.~\ref{case2}(d). Here the top  cords of $R_p$ cannot reach their corresponding corners, which is fine since the adjacencies  $R_{b'},R_p$ and $R_{b},R_p$ are realized at the top-right and bottom-right cords of $R_p$, and since for each rectangle adjacent to the top of $R_p$, one of its two cords is extended to touch the cords of $R_p$. 
 Furthermore, we will never need to choose between the top arms 
 of $R_p$ to process the remaining consumer rectangles.
 
 \item Case ($\delta_b> d_p /2$):

 In this case, we will always saturate the bottom-right cord.
If $R_{b'},R_b\subseteq_y R_p$, then we follow Fig.~\ref{case2}(e). 
If $R_p\subset_y R_b$ and $R_{b'} \subseteq_y R_p$, then we follow  Fig.~\ref{case2}(f) or (g) depending on the placement of $R_g$. 
The case when $R_p\subset_y R_{b'}$ and $R_b\subseteq_y R_p$ is symmetric.  
Finally, if   $R_p\subset_y R_b$ and $R_p\subset_y R_b$, then we follow   Fig.~\ref{case2}(h). Here the right  cords of $R_p$ cannot reach their corresponding corners, which is fine since the adjacencies  $R_{q},R_p$ and $R_{b},R_p$ are realized at the top-right and top-left cords of $R_p$, and since for each rectangle adjacent to the bottom of $R_p$, one of its two cords is extended to touch the cords of $R_p$.

 \item Case ($\delta_l> d_p /2$):

 In this case, we will always saturate the bottom-left cord.
 If $R_b\subseteq_y R_p$, then we follow Fig.~\ref{case2}(i). 
If $R_p\subset_y R_b$, then we follow  Fig.~\ref{case2}(j) or (k) depending on the placement of $R_g$.
%
In the latter case, the top-right and bottom-left  cords of $R_p$ cannot reach their corresponding corners, which is fine since the adjacencies  $R_{q'},R_p$ and $R_{b},R_p$ are realized at the bottom-right and top-left cords of $R_p$, and since for each rectangle adjacent to the left of $R_p$, one of its two cords is extended to touch the cords of $R_p$.

 \item Case ($\delta_r> d_p /2$):

 This case is symmetric to the case when $\delta_l> d_p /2$.

 \item Case ($\delta_l,\delta_r,\delta_t,\delta_b\le  d_p /2$):

 This case can be handled in the same way as the case when $\delta_t > d_p /2$, i.e., following the Fig.~\ref{case2}(b)--(d).
We cannot always saturate a cord now, but by suitably splitting $\delta_t$ between two cords, we can ensure that all cords are balanced.
\end{itemize}


\end{appendix}

\end{document}